# Wrangling Data Issues to be Wrangled: Literature Review, Taxonomy, and Industry Case Study

QIAOLIN QIN, Polytechnique Montreal, Canada
HENG LI, Polytechnique Montreal, Canada
ETTORE MERLO, Polytechnique Montreal, Canada

Data quality is vital for user experience in products reliant on data. As solutions for data quality problems, researchers have developed various taxonomies for different types of issues. However, although some of the existing taxonomies are near-comprehensive, the over-complexity has limited their actionability in data issue solution development. Hence, recent researchers issued new sets of data issue categories that are more concise for better usability. Although more concise, modern data issue labeling's over-catering to the solution systems may sometimes cause the taxonomy to be not mutually exclusive. Consequently, different categories sometimes overlap in determining the issue types, or the same categories share different definitions across research. This hinders solution development and confounds issue detection. Therefore, based on observations from a literature review and feedback from our industry partner, we propose a comprehensive taxonomy of data quality issues from two distinct dimensions: the *attribute* dimension represents the intrinsic characteristics and the *outcome* dimension that indicates the manifestation of the issues. With the categories redefined, we labeled the reported data issues in our industry partner's data warehouse. The labeled issues provide us with a general idea of the distributions of each type of problem and which types of issues require the most effort and care to deal with. Our work aims to address a widely generalizable taxonomy rule in modern data quality issue engineering and helps practitioners and researchers understand their data issues and estimate the efforts required for issue fixing.

CCS Concepts: • **Information systems** → **Data cleaning**.

Additional Key Words and Phrases: Data Quality, Data Issue Detection, Data Quality Issue Taxonomy

## 1 INTRODUCTION

Data play important roles in modern society; for example, data are essential for modern software products, services, and research [5, 7, 16, 42]. For software or other products or services built based on data, the data quality is critical to the users' perceptions of the quality of these products and services [4, 18]. A consensus reached on data quality is that high-quality data facilitate downstream projects; conversely, poor-quality data may lead to people paying high costs when applying them [30] [17]. For example, researchers have raised concerns about the negative influence of the insufficient quality of software engineering data [9, 23, 32, 41]. As a consequence, it is essential for researchers and practitioners to understand their data, detect data quality issues, and fix data problems.

Four steps were stated by Wang *et al.* to manage data quality, namely **define**, **measure**, **analyze**, and **improve** [39]. This management loop highlights the importance of data quality constraint definition, which serves as a basis for the following management actions: with a high-quality constraint set specified, data analysts can use tools to detect data issues that violate the constraints,

Authors' addresses: Qiaolin Qin, qiaolin.qin@polymtl.com, Polytechnique Montreal, Montreal, Canada; Heng Li, heng.li@polymtl.com, Polytechnique Montreal, Montreal, Canada; Ettore Merlo, ettore.merlo@polymtl.ca, Polytechnique Montreal, Montreal, Canada.

search for fixing strategies that fit certain issue types, measure current database quality, and analyze the histories for each category to improve their operation loop; the researchers, on the other hand, can refine their issue-detecting or fixing models for specific categories.

As the elements to be measured and fixed, data quality issues arise from violation to quality constraint rules [10]. According to Wang *et al.*, data quality studies can be categorized into three types to meet different needs [40]: **intuitive** (*i.e.,* results are derived from experience or intuitive comprehensions [27] [3]), **theoretical** (*i.e.,* focus on the data deficiency during the data manufacturing process [37]), and **empirical** (*i.e.,* estimates whether the data are suitable for consumers to use [20]). Taxonomy sets for data quality issues previously defined using different study strategies on various dimensions are near-comprehensive, and greatly facilitate the traditional data issue detection algorithms: Rahm *et al.* categorized the data issues based on their sources [29], Kim *et al.* used a tree structure, which is "near-complete", to pose a workflow in determining the issue types [21]; Oliveira *et al.* removed the existence level separation in the previous research [29], and fine-grained the taxonomy using relationships [26]. However, the over-details of the previous taxonomies are placing an obstacle in modern use: for example, the tree-structured categorization provided by Kim *et al.* [21] grows the deepest branch into 6 layers, indicating its laborious while labeling.

We notice that previous categorization rules are unsuitable for wrangling issues in big data due to its excessive workload. Conversely, the quality constraints being discussed are implicitly merged into **data quality issue detectors and cleaners**. Instead of leveraging detailed rules to measure data quality and refine the database manually, this modern solutions [2] [19] [25] [24] [22] intuitively cluster data quality issues by their nature and are established from data perspective [13] (*e.g.,* RAHA combines multiple data-issue detecting tools and captures four types of issues according to their threat to the database, namely outliers, pattern violations, rule violations and knowledge base violations [25]). These solutions are mostly modularized based on issue clusters, providing convenience when users try to fine-tune certain detection or fixing components. Although multiple data quality solution systems are developed, the field lacks a comprehensive and mutually exclusive taxonomy set for modern issue categories, and the terminologies used in different research vary: Abedjan *et al.* assign data issues to the four main types of issue detectors based on the constraint they violate (*e.g.,* outliers, duplicates, rule violations, and pattern violations) [2]; Mahdavi *et al.* replaced the previous duplicate issue with knowledge base violations to cover a boarder range of data issues [25]; instead of deriving from the four data issue families, Hynes *et al.* split the data issues into three types, namely miscoding, outliers and scaling, and packaging error [19]. Problems caused by inadequate definition on modern quality issue categorization include:

- Caused by weak boundary, **issue attributes are not precisely defined**. (*e.g.,* the difference between integrity issues and data smells addressed by Foidl *et al.* is whether the issue is *observable*, while the data smells may also be observable when the data is validated in a human-in-loop fashion [12].)
- Caused by lack of a detailed categorization workflow, **terminology definitions overlap within a study** [36]. (*e.g.,* according to Abedjan *et al.*, the pattern violations indicate conflicts to syntactic and semantic constraints, which overlaps with integrity constraints indicated in rule violations [2].)
- Caused by inadequate analysis, **the taxonomy set does not cover all the issue types**. (*e.g.,* although the data are being validated on different dimensions by Lwakatare*et al.*, the issue types addressed in the literature do not cover inter-column violations [24].)
- Caused by ambiguous definition, **the terminologies within a taxonomy set are not defined on the same dimension**. (*e.g.,* in the four data quality issue families provided by

  Abedjan *et al.*, duplicates are observable issue attributes, while rule violations and pattern violations explicitly state the issue impact on databases [2].)
- Caused by taxonomy misalignment across studies, **the labels are not generalizable and thus cannot be used for system comparison**. (*e.g.,* the definition of rule violations by Abedjan *et al.* indicates any type of violation to integrity constraints [2], while according to Mahdavi *et al.*, rule violation only indicates the issues concerning multiple columns [25].)

The problems in issue category sets will lead to confusion for future researchers and practitioners when they design or refine data issue-detecting or fixing systems. Moreover, without a consensual and actionable labeling system, practitioners are not able to compare detector performances in parallel, measure data quality, and improve the database using suitable issue-fixing strategies.

Our industry partner (*CompanyX*) has a large-scale data warehouse (*DW-X*) that stores information of millions of users across the globe and is updated on a daily basis. The data are provided by multiple data suppliers (*e.g.,* websites) and consumed by multiple consumers (*e.g.,* search services). Due to the heterogeneous nature of the data, the data warehouse suffers from an overwhelming number of data issues, many of which are reported by the data consumers, causing a significant negative impact on the industry partner's productivity and quality of service. To suppress the negative impact data quality issues have on the downstream products, we aim to design a data quality engineering framework that can help developers of data-centric systems efficiently detect, classify, and address data-related issues. Such a framework requires parallel comparison among existing data issue solutions during development, which necessitates a taxonomy of data-related issues. However, we could not find a satisfactory taxonomy after investigating the literature.

Given the need for data quality issue taxonomy in researching modern data quality solutions, we derived a set of issue categorization rules according to recent studies and industrial experiences. Following the instructions provided by Kim *et al.* [21], we designed tree-structured labeling workflows to facilitate future use. With the taxonomies defined, we label a sample of Jira issue tickets in a real-world commercial company data warehouse (*DW-X*). In comparison to manually created toy datasets, real-world datasets are more accurate in reflecting real distributions [34]. Hence, the comprehension built and solutions designed based on these datasets have a higher reliability when transferred to other problems. The labels obtained from the data warehouse allow us to establish a statistical understanding of the data issue distribution in real-world scenarios. Further, according to the ticket information, we further analyze the difficulty level in handling each category of issue. The understanding gained from the data issues of the data warehouse can guide practitioners in improving database management procedures and finding corresponding fixing strategies.

Our work makes the following main contributions:

- We provide a review of the studies on data quality issues in the past 10 years and compares the data quality issue categories determined in these studies.
- We present a comprehensive and novel taxonomy of data quality issues, including a hierarchical set of mutually exclusive data issue categories covering integrity-breaking issues and data smells.
- In our proposed taxonomy, we categorize data quality issues from two distinct dimensions: (1) an *attribute* dimension that captures the intrinsic characteristics of the issues (i.e., their semantic impact on the database), and (2) an *outcome* dimension that captures the observable manifestation of the issues which can suggest concrete detection solutions.
- We present an analysis of the presence of the data issue categories in a real-world, large-scale data warehouse, to provide an insight into the practical significance of these data quality issues.

- We discover which types of data issues are the most difficult for developers to fix through an analysis of detailed ticket information recorded on Jira, to provide insight for developers to manage and prioritize their data quality issues.

**Organization.** The remainder of our study is organized as follows. Section 2 introduces our literature review on data quality studies in the past 10 years. Section 3 addresses two sets of taxonomies and their labeling process on the outcome dimension and attribute dimension; the two sets of taxonomies are bridged up to aid researchers and practitioners in understanding data quality issues from two perspectives. Section 4 applies the rules previously defined on a real-world data warehouse issue set and observes the distributions and difficulty levels of fixing each type of issue. Section 5 discusses the validity threats we may face during the research. Section 6 provides an insight into previous research related to our work. Finally, Section 7 concludes the paper

## 2 A LITERATURE REVIEW ON DATA QUALITY ISSUES

To carry out the literature review, we first collected an initial set of literature on data quality issues with a Google Scholar search [25][22][12]. Based on the initial literature, we implemented a snowballing method (including both forward and backward snowballing) to collect more relevant studies.

According to the definition given by Ge and Helfert [13], data quality issues can be categorized by their relevance to the context (*i.e.,* a context-dependent issue indicates that the issue violates domain specifications, while a context-independent issue suggests that the issue can be generalized to databases in other domains) [13]. Inspired by risk-based data validation [11], Foidl *et al.* [12] bring the context-independent issues to a finer granularity by their level of suspiciousness. According to Foidl *et al.*, issues with higher suspiciousness are easier to catch by traditional techniques and, thus, are **obvious** to developers; the less suspicious issues have uncertain effects on data and are hard to filter, therefore, are being categorized as **latent** issues. Inspired by the term "code smells"[1], Foidl *et al.* addressed the latent context-independent issues as **data smells**.

However, the determination of *obvious* and *latent* in Foidl *et al.* is somehow unclear in practice: a data smell could be latent for a detecting system while being obviously detected and fixed if a system was designed human-in-loop. For further clarification of data smells, we revisit the definition of code smell: code smells are **symptoms of poor design and implementation choices** [35]. Different from bugs, code smells do not directly break the integrity of the project. The project's integrity-friendliness indicates the legitimacy of the code smells; the bugs would prevent the project from functioning and, thus, are not legal.

To align the term with code smells, data smells are defined as arising from poor data handling in databases [31] [14]. We further determine that data smells are legal issues that **do not result in breaking the integrity of the database directly**. Conversely, the issues that directly break database integrity share more of the same attributes as the bugs in a project, which are named integrity issues in the following contents. Therefore, we classify the data issues into two broad categories: **integrity issues** and **data smells**. Our literature review is implemented along the two categories.

- *Integrity issues*: Data issues that directly break the database's integrity (e.g., duplicate records).
- *Data smells*: Data issues that do not directly break the database's integrity but that may negatively impact the maintenance and usability of the database (e.g., inconsistent records).

One of our main aims is to align the categories among modern data quality issue solutions. Therefore, we are to collect the data issue labels and the definitions for each label in previous research. The related literature is collected following the following criteria:

- All the literature should be published within the past 10 years (*i.e.,* 2013 to 2023) to avoid being outdated.
- The literature should be primarily related to data quality issues (either integrity issues or data smells).
- For integrity issues, the literature should propose a taxonomy of data quality issues or a detection solution for multiple types of data quality issues. It should contain explicit introductions for issue types: the explicit definition shall allow us to avoid a biased understanding of terminology, and thus, they can be compared horizontally to align with other research.
- For data smells, as the terminology is novel in research, the literature should explicitly define and describe the terminology "data smells" and potential sub-categories to facilitate understanding.

Our literature review identified 12 papers related to data quality issues, including 8 for integrity issues and 4 for data smells. According to our review, there are no studies comparing integrity issues and data smells in detail. Therefore, we address studies related to the two types of issues separately in the following paragraphs.

## 2.1 Integrity Issues

Table 1 provides an overview of the recent research related to data integrity issues. We collected 8 papers that explicitly define the issue categories covered by their data quality issue solutions.

Table 1. Literature related to data integrity issues.

| Literature | Year | Categories | Definitions |
| --- | --- | --- | --- |
| Detecting Data Errors: Where Are We and What Needs to be Done?[2] | 2016 | Outliers | Data values that deviate from the distribution of values in a column of a table. |
|  |  | Duplicates | Distinct records that refer to the same real-world entity. |
|  |  | Rule violations | Values that violate any kind of integrity constraints. |
|  |  | Pattern violations | Values that violate syntactic and semantic constraints. |
| The Data Linter: Lightweight, Automated Sanity Checking for ML Data Sets[19] | 2017 | Miscoding | Data that should be transformed to improve the likelihood that a model can learn from the data. |
|  |  | Outliers and Scaling | The data contains uncommon list length, uncommon signs, unformalized features, or significantly extreme values. |
|  |  | Packaging Error | Problems with the organization of the data. |
| Metadata-Driven Error Detection[36] | 2018 | Pattern violations | Syntactic pattern violations, misspelled linguistic data, formatting rule violations, or wrong semantic data type affiliation. |
|  |  | Rule violations | Inconsistencies and conflicts according to rules or integrity constraints. |
| | | Outliers | Data points that significantly deviate from the distribution or structure of the remaining data points. |
| | | Duplicates | Tuples that refer to the same real-world entity. |
| Uni-Detect: A Unified Approach to Automated Error Detection in Tables[38] | 2019 | Numeric outliers | Values that deviate significantly from the underlying distribution. |
| | | Spelling errors | (*e.g.,* "Mississippi"and "Missisipi") |
| | | Uniqueness constraint violations | Certain columns are semantically required to be unique (*e.g.,* an ID column), where duplicate values can be flagged as errors. |
| | | Functional-dependency (FD) violations | FD violations are conceptually similar to Uniqueness, but are defined over two groups of columns. |
| Raha: A Configuration-Free Error Detection System[25] | 2019 | Outliers | (Violate) the general distribution of values that reside inside the column. |
| | | Pattern violations | Data values that do not match a certain pattern. |
| | | Rule violations | (Violate) rules in the form of functional dependencies (FDs). |
| | | Knowledge base violations | Data values in the matched data columns that conflict the entity relationship inside the knowledge base. |
| On the Experiences of Adopting Automated Data Validation in an Industrial Machine Learning Project[24] | 2021 | Outliers | |
| | | Duplicates | Redundant values |
| | | New/missing features from baseline | Features present. |
| | | Not in range | Features (does not) have right values. |
| CleanML: A Study for Evaluating the Impact of Data Cleaning on ML Classification Tasks[22] | 2021 | Missing values | No value is stored for cells. |
| | | Outliers | An outlier is an observation that is distant from others. |
| | | Duplicates | Duplicates refer to the records that correspond to the identical real-world entity. |
| | | Inconsistencies | Inconsistencies occur when two cells in a column have different values, but should actually have the same value. |
| | | Mislabels | Mislabels occur when an example is incorrectly labeled. |
| Automated Data Cleaning Can Hurt Fairness in Machine Learning-based Decision Making[15] | 2023 | Missing values | NULL and NaN values in the datasets. |

| | |
|---|---|
| Outliers | Data more than n standard deviations away from the mean of the column or lies outside of the interval between the 75th and 25th percentile; tuple identified as such by an isolation forest trained on the data. |
| Label errors | Tuples with the wrong prediction label assigned to them. |

As shown in Table 1, existing taxonomies have proposed different sets of data issue categories. These taxonomy sets may only cover a small number of issue categories (*e.g.,* no inter-column issue is being addressed by Lwakatare *et al.* [24]) or categories that overlap with each other (*e.g.,* rule violations and duplicate provided by Abedjan *et al.* [2] overlaps with each other, since the unique constraint is also a kind of integrity constraint). Moreover, the terminologies used for integrity issues with similar definitions vary across studies (*e.g.,* the term "miscoding" in data linter [19] refers to the pattern problems in the database, which are indicated as "pattern violations" in the other two papers [2][25]). Based on our observation of the categories, we found that there exist two types of terminologies: the first type addresses certain constraints that the issue violates (*e.g.,* pattern violations arise when a piece of data conflicts with given pattern constraints), while the second type is concerned with the attributes of the issues (*e.g.,* duplicate does not describe the violation to constraints directly, but it exists as an intrinsic characteristic of being "not unique"). The two types of terminologies describing the data quality issues from different perspectives were mixed in the previous studies (*e.g.,* Abedjan *et al.* leverage the category *duplicates* together with *rule violations* and *pattern violations* [2]). The unaligned terminologies and differences between their dimension of definition would cause a problem when assigning labels for the data issues, which could further affect the quality of solution assignment and system development. To avoid dimension mixture, it is essential to distinguish the two types of categories when designing the taxonomies.

## 2.2 Data Smells

The definition of data smells is derived from the notion of code smells [31] [12]. Contradicting the integrity problems, data smells do not break the integrity of a database and are hard to handle. Although the separation between integrity issues and data smells based on observability is intuitive, it can barely serve as a strong boundary; only a few studies on data smells contain fine-grained sub-categories with descriptions. Since the terminology is not being widely researched, we only collect 4 papers (shown in Table 2) with definitions for *data smells* sub-categories.

The sub-categories of data smell proposed in existing taxonomies are derived from the observation of databases. Based on previous analyses, the basic term *data smells* has the feature of not causing conflicting outcomes. Provided with this feature, sub-categories of data smells cannot be further split on the outcome dimension but only on the attribute dimension since they do not directly conflict with the constraints.

## 3 A COMPREHENSIVE DATA QUALITY ISSUE TAXONOMY

In this section, we define the data quality issue taxonomies on two dimensions: *attribute* and *outcome*. The design integrates feedback from our industrial partner, who seeks a taxonomy system to document previous data quality issues, in the following fashion:

**Step 1:** Derive an initial set of quality issue taxonomies from the literature review.

Table 2. Data smells literature review.

| Literature | Year | Categories | Definitions |
|---|---|---|---|
| Risk-Based Data Validation in Machine Learning-Based Software Systems [11] | 2019 | Miscoding smells | *e.g.* Date/time or numbers encoded as a string |
| | | Categorical value smells | *e.g.* Values of input data signals vary widely. |
| | | Intermingled Data Types | *e.g.* Using 'O' instead of '0'. |
| Data Smells in Public Datasets [33] | 2022 | Redundant value smells | Smells which occur due to presence of features that do not contribute any new information. |
| | | Categorical value smells | Smells which occur due to presence of features containing categorical data. |
| | | Missing value smells | Smells which occur due to absence of values in a dataset. |
| | | String value smells | Smells which occur due to presence of features containing string type data. |
| Data Smells: Categories, Causes and Consequences, and Detection of Suspicious Data in AI-based Systems [12] | 2022 | Believability smells | Semantically implausible data values. |
| | | Understandability smells | The inappropriate, unusual or ambiguous use of characters, formats or data types. |
| | | Consistency smells | The use of inconsistent syntax with respect to data values in partitions of data. |
| Preliminary Findings on the Occurrence and Causes of Data Smells in a Real-World Business Travel Data Processing Pipeline [14] | 2022 | Encoding smells | Encoding smells describe the problems connected with inappropriate data types. |
| | | Consistency smells | The same data have nonidentical expressions. |
| | | Syntactic smells | Syntactic smells represent inappropriate expressions of values that might lead to misinterpretation of information by humans or algorithms. |
| | | Believability smells | Believability smells can be interpreted by software as correct data and understood by humans, but they do not represent the right data. |

**Step 2:** Introduce the definition of taxonomies to the data analysis team at *CompanyX*. The data scientists are then required to individually label identified data issues in several tables from their data warehouse.

**Step 3:** Gain feedback from the data analysis team by discussing on the labeling results.

**Step 4:** Refine the data quality issue taxonomies based on the feedback.

Our taxonomies aim to capture the features of the quality issues from different perspectives while solving the dimension mixture problem stated in 2.1. The taxonomy of the attribute described in subsection 3.1 is based on the characteristics of the issues (*e.g.,* a negative number in the "age" column can be described as "data invalid" since valid values shall be non-negative), while the taxonomy of the outcome dimension described in subsection 3.2 is based on the constraint an issue breaks (*e.g.,* a negative number in the "age" column triggers "range violation", that the valid range shall be *age* >= 0). In subsection 3.3, we align the definitions on the two dimensions to facilitate future use (*e.g.,* data issue labeling). For better understanding, we set up a simple scenario of **a database at a convenience store** to demonstrate each type of data quality issue.

> **A Convenience Store Example**
>
> A convenience store establishes a database with two tables to record its products and customer purchase records. The product table includes fields such as *ProductID, ProductName, ProductPrice, Discount, FinalPrice*; the purchase record table includes fields such as *PurchaseID, CustomerID, ProductIDList, PurchaseTotal*. The management constraints are as follows:
> (1) Each product shall have a *ProductName* and a non-negative float type *ProductPrice* as the original price.
> (2) To ensure the uniqueness of products, the *ProductID*s are non-duplicatable 5-digit integers.
> (3) If the product is on sale, the *Discount* would be larger than 0 while not larger than *ProductPrice*, and the *FinalPrice* equals to a *ProductPrice - Discount*. On the other hand, if the product is not on sale, the *Discount* should remain 0, and the *FinalPrice* equals to the *ProductPrice*.
> (4) Each purchase record is unique, and the *PurchaseID* should be non-duplicatable 10-digit integers.
> (5) If the customer recorded his/her *CustomerID* during checkout, a 8-digit *CustomerID* would be recorded, otherwise the *CustomerID* should be inapplicable.
> (6) The *ProductIDList* is a duplicable list, containing the IDs of all the products purchased in the record. All the *ProductID* in *ProductIDList* shall be registered in the product table. The *PurchaseTotal* equals the sum of *FinalPrice*s for every purchased product in the list.

**3.1 Taxonomy on the Attribute Dimension**

In this subsection, we bring integrity problem categorization onto the dimension of attributes based on our literature review. Different from the outcome dimension, the attributes of a data quality issue do not explicitly address the constraint violation but describe the features of the issue. Fig. 1 provides an overview of the issue taxonomy of the dimension of attributes. The tree-structured plot can be used to guide the process of determining the label for an issue.

*3.1.1 Integrity Issues.* The integrity issues have the attribute of being illegal in the database. To simplify the categorical representation, we mimic the term definitions in wireless communications and create a list of concerns for columns and rows. The abbreviations used are listed as follows:
- **SCSR**: Single column, single row.
- **SCMR**: Single column, multiple rows.
- **MCSR**: Multiple columns, single row.
- **MCMR**: Multiple columns, multiple rows.

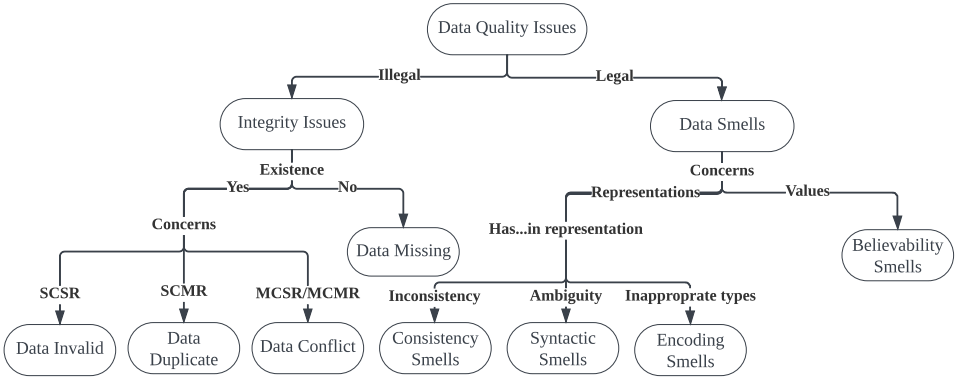

Fig. 1. An overview of the data quality issue taxonomies on the dimension of attributes.

**Data Missing/NULL**

The reason for a piece of data not existing in the database or a database containing "NULL" valuemay vary, but both problems indicate a knowledge piece that should exist but is lost in the real database.

> **Data Missing Example**
>
> - Cell Data Missing
>   The *ProductID* is missing for one product in the product table.
> - Inter-Row Data Missing
>   When checking the purchase record table after closing the store, the manager found that the last purchase record of today, which should be documented a row, was missing.
> - Inter-Table Data Missing
>   According to the *ProductIDList* in a purchase record, a customer bought a product whose *ProductID* cannot be found in the product table, indicating the product is missing in the product table.

**Data Invalid**

Data invalid can sometimes be mingled with NULL data, according to the experience of our data scientists. Hence, under our taxonomy rules, data invalid occurs only when the information piece exists. A piece of invalid data conflicts with predefined domain rules and can be checked solely based on single-column-single-row(SCSR), without comparison with other rows or columns.

> **Data Invalid Example**
>
> According to the second constraint stated in the example scenario, a *ProductID* should be a 5-digit integer; however, one of the *ProductID*s in the product table is observed as a 4-digit integer.

**Data Duplicate**

Data duplication can only happen among multiple rows. To determine whether the rows are duplications, a single key shall be selected to identify the same entities. Thus, the duplicate data problem is concerning single-column-multiple-row(SCMR).

> **Data Duplicate Example**
>
> Two product records with the same unique indicator (*i.e.,* ProductID) and other same information (*i.e., ProductName, ProductPrice, Discount, FinalPrice*) exist in the product table.

**Data Conflict**

Data conflict exists in either columns or rows. The column conflict concerns multiple-column-single-row (MCSR), while the row conflict concerns multiple-column-multiple-row (MCMR). Note that new/missing features from the baseline [24] and mislabels [22]/label errors [15] also belong to this category, since they are conflicts with baselines. The following two examples demonstrate the two types (*i.e.,* MCSR and MCMR) of conflicts:

> **Data Conflict Examples**
>
> - Inter-Column Conflict
>   A product on discount has the *ProductPrice* as 10.00, and the *Discount* as 2.00, but its *FinalPrice* equals 7.50 instead of 8.00. The three different columns (*i.e., ProductPrice, Discount, and FinalPrice*) conflict with the third constraint.
> - Inter-Row Conflict
>   Two product records with the same unique indicator (*i.e.,* ProductID) are not duplicates but have different *FinalPrice* in the product table, causing a conflict.
> - Inter-Table Conflict
>   A customer purchased two products whose *FinalPrice*s in the product table are 5.00 and 2.00, respectively. However, the corresponding row in the purchase record table indicates that the *PurchaseTotal* is 9.00 instead of 7.00. The inequality violates the sixth constraint.

*3.1.2 Data Smells.* Compared with the previous study by Foidl *et al.* [12], the category of data smells was brought to a finer granularity in the research of Golendukhina *et al.* [14] (*i.e.,* the previous understandability smell is decomposed into syntactic smells and believability smells with a detailed definition). Based on the attributes, data smells are categorized into four types.

**Believability Smells**

Believability smells refer to data that is highly suspicious in value but interpretable in semantics. Given the definition of suspiciousness, believability smells also include outliers from a data distribution: the outliers are not illegal to a dataset. Still, they are to be suspected due to their low likelihood of existence.

> **Believability Smells Example**
>
> According to the price calculation rule, an acceptable *FinalPrice* shall be a non-negative float number. Hence, 0.0 shall be a valid *FinalPrice* (*i.e.,* representing the product can be taken for free). However, it could be suspicious that 50% of the products have 0.0 as their *FinalPrice*.

**Consistency Smells**

By stating consistency smells, we cover the legal issues for stating the same entity with inconsistent representation. More specifically, consistent smells arise from different representations of the same entity and could cause problems in future data handling.

> **Consistency Smells Example**
>
> Several rows in the purchase record table have inapplicable *CustomerID*, as some customers do not provide their membership card at the cashier. Some inapplicable *CustomerID*s were recorded as "NULL" while the others were NaN. The different representations of information absence could cause future problems when processing the table.

**Syntactic Smells**

According to Golendukhina *et al.*, syntactic smells are inappropriate expressions of values that result in ambiguity. This type of smell shares some similarities with consistency smells, since they both concern data expressions. Although both concern data representations, syntactic smells refer to the ambiguity in the representation of a **single data unit**, while consistency smells compare across **multiple data units** that refer to the same entity with multiple representation types.

> **Syntactic Smells Example**
>
> There are two types of apples for sale at the convenience store. The two types of apples have different *ProductID* and *ProductPrice*, but both have the same *ProductName* as "Apple". Ambiguity could arise when products are wrangled by their *ProductName*s.

**Encoding Smells**

Encoding smells are created by misusing characters, formats, or data types. This type of smell also addresses issues in data representations in a single data unit but does not cause ambiguity in representation; encoding smells is a symptom of not representing data with proper data types.

> **Encoding Smells Example**
>
> According to the manager's experience, *ProductName* shall always be a string. Thus, downstream programs process *ProductName*s as a string type record by default. However, since there is no constraint on *ProductName* existing in the table, a robot toy with *ProductName* "0010010" was encoded as an integer 10010. The encoding smell causes problems on downstream programs since they are not being coded to deal with integer data in this field.

### 3.2 Taxonomy on the Outcome Dimension

Solutions to different types of data corruption may vary. To facilitate the future use of researchers and data scientists in understanding, detecting, documenting, and fixing the manifestation a data issue would bring to the database, we provide a set of mutually exclusive categories on the outcome dimension along with their definitions and the corresponding label determination workflow. Fig. 2 illustrates an overview of the issue types derived from recent data quality issue studies.

*3.2.1 Integrity Issues.* The research by Mahdav *et al.* extends the categories mentioned in a previous research [2] and defines four types of problems (*i.e.,* outliers, pattern violation, rule violations and knowledge base violations) based on the issue's position of existence [25]. With the layered definition strategy, the categories covered majority of the issues. As a refinement, we derive our definition of four integrity problem families based on the ones defined in the research [25].

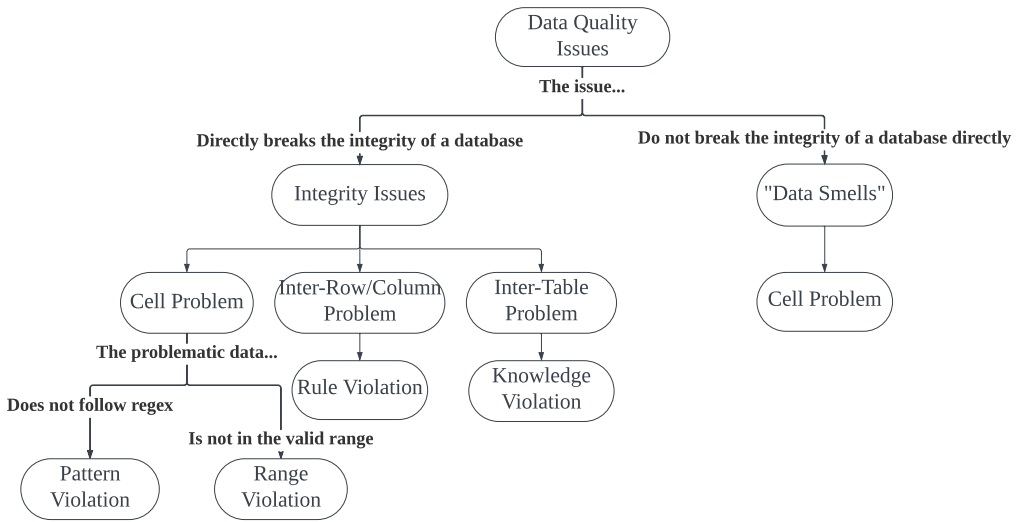

Fig. 2. An overview of the data quality issue taxonomies on the dimension of outcomes.

**Pattern Violation**

Pattern violation detection systems check the validity of data values according to a set of predefined data patterns. The values in each row are checked by column; thus, unit pattern violations exist only within cells. By stating *pattern*, we indicate syntactic constraints that are the restrictions on forms and types of values. Pattern violations are conflicts with fixed regular expressions or data types.

> Pattern Violation Example
>
> Similar with the example provided in **data invalid**: A *ProductID* in the product table is observed as a 4-digit integer, whereas it should be a 5-digit integer.

**Range Violation**

We substitute the term *outlier* in previous research [25] with "range violation", since the previous term indicates more of an observed anomaly rather than an explicit conflict with predefined database rules: given the subjective and threshold-dependent nature of "suspiciousness", the observed anomaly may have a high level of suspiciousness according to a badly determined manual threshold but can still be legal to the database and is integrity-friendly. The terminology is derived from *not in range* [24], but highlights the conflict with range constraints. Similar to pattern violation, range violations exist within the unit of cells. This type of violation can be captured through comparisons between values and a valid range list (*i.e.,* a predefined range of accepted values). The range can be of any data type and can be either finite or infinite, while with a set of fixed constraints.

> **Range Violation Example**
>
> According to the third constraint, the *Discount* should be a non-negative value; however, one of the products has a negative *Discount*.

**Rule Violation**

According to Mahdav *et al.*, single column problems (*i.e.,* cell problems) can be covered by **pattern violations** and **outliers**. In contrast, rule violation only refers to inter-column data issues, also defined as a conflict with functional dependencies. The previous definition thoroughly considered the relationship between columns but neglected the relationship between rows. As an extension, we promote rule violations to be both inter-columns and inter-rows.

> **Rule Violation Example**
>
> - Inter-Column Rule Violation
>   The same example with inter-column **data conflict**: a product on discount has the *ProductPrice* as 10.00, and the *Discount* as 2.00, but its *FinalPrice* equals 7.50 instead of 8.00.
> - Inter-Row Rule Violation
>   According to the fourth constraint, purchase records should be unique. However, two rows in the purchase record table are duplicates.

**Knowledge Violation**

Existing research such as Katara[8] define knowledge violations as issues detected by leveraging open-source knowledge bases. The open-source knowledge bases are usually robust in general knowledge and are feasible to use; however, they may lack knowledge in specific fields or private knowledge. The knowledge in particular areas or is private may be stored in other types of repositories instead of knowledge bases(*e.g.,* in different tables of the database), and according to existing definitions, violations of this type of knowledge are not being covered by either knowledge base violation or other problem types. Hence, we extend the knowledge base violation to knowledge violation to represent all the inter-table problems: the search and comparison for the knowledge are made inter-table (*e.g.,* the problem in one of the tables is detected by verifying with knowledge in another table).

> **Knowledge Violation Example**
>
> The same example with inter-table **data conflict**: A customer purchased two products whose *FinalPrice*s in the product table are 5.00 and 2.00, respectively. However, the corresponding row in the purchase record table indicates that the *PurchaseTotal* is 9.00 instead of 7.00, causing a conflict across tables.

*3.2.2 Data Smells.* Derived from code smells, the term data smells concerns data issues introduced to databases by poor design and do not directly break the integrity constraints. According to the literature, data smells are limited to the dimension of cells. Since data smells do not violate any constraint directly, these issues cannot be divided into finer granularity on the outcome dimension. The difference between data smells and range violation problems is that the data smells are not restricted to a predefined valid range and rise from the poor design of constraints.

## 3.3 Category Alignment for Integrity Issues

To better measure, analyze, and improve the database [39], we leverage the two dimensions addressed in Sec 3.1 and Sec 3.2 to obtain a better comprehension of the integrity issues. Since data smells do not cause a direct violation of the database constraints, this type of issue is only defined into sub-categories on the attribute dimension. Consequently, data smells do not require category alignment. Integrity issues, conversely, have distinct attributes that could cause a direct impact on the database: the two dimensions being determined in the previous subsections aim to characterize the data quality issues from two different perspectives (*i.e.,* issue attribute and its defective outcome). According to the characteristics of the integrity issues, the taxonomies for integrity issues on the two dimensions can be aligned through a set of rules, which are presented in Fig 3.

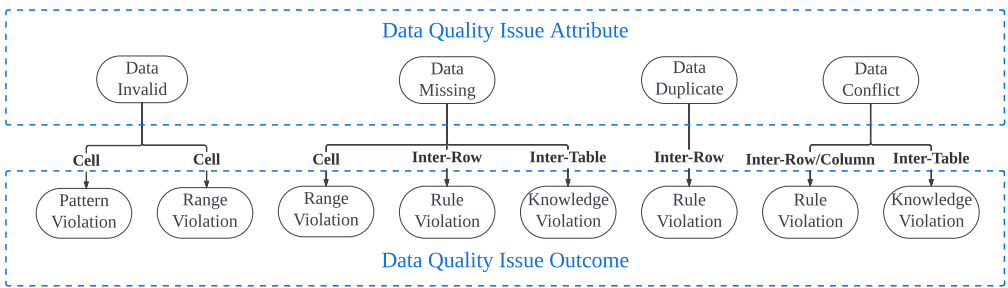

Fig. 3. The alignment of categories on the two dimensions.

Fig 3 provides a set of rules for determining the defective outcome caused by issues with different types of attributes. The linkage is provided through the existence position of the data issue (*i.e.,* cell, inter-row, inter-column, inter-table). Through the determination rules, we cover the data quality issues on different scopes (*e.g.,* according to the convenience store example, when a product on discount has the ProductPrice as 10.00, and the Discount as 2.00, but its FinalPrice equals 7.50 instead of 8.00, the issue is a data conflict while it causes rule violation). These taxonomies can serve as guides for researchers and practitioners in data quality issue labeling, detector fine-tuning, and repair system designing.

**Summary**: Based on a review of existing studies on data quality issues, we observe that data quality issues can be characterized from two distinct dimensions: the intrinsic characteristics of the issues (e.g., data duplicate) – the attribute dimension, and the manifestation of the issues in terms of constraint violations (e.g., rule violations) – the outcome dimension. We thereby propose a comprehensive taxonomy of data quality issues with two hierarchical structures to cover the two dimensions.

## 4  A CASE STUDY OF DATA QUALITY ISSUES IN A LARGE-SCALE DATA WAREHOUSE

In this section, based on the taxonomy derived from the literature review (Section 2), we examine a set of data quality issues in 3 tables from a real-world large-scale data warehouse (*DW-X*). The set of issues is provided by a large private company (*CompanyX*) with millions of clients; the quality of the data is critical to the company as the data are used by multiple consumers (e.g., customer relationship programs) to provide high-quality services to the clients.

In total, we examined 94 data issue tickets reported related to 3 large product tables from *DW-X* in the JIRA issue management system of *CompanyX*. The tickets contain information such as the severity, priority, detailed description, etc., for each data issue. We collaborate with the data analysis team, who understand the data well and can fully access to the data warehouse at *CompanyX* to label the data issues on both the outcome and attribute dimensions. With the issue labels obtained, we observe the distribution of the data issues and estimate the level of difficulty in dealing with each type of data issue.

*Labelling process:* During data quality issue labeling, we first provided the taxonomies together with definitions to the data analysis team. Each class has an example from *CompanyX*'s JIRA tickets to aid our data scientists' understanding. To provide consensual and unbiased understanding, each ticket is initially labeled individually by at least 2 persons. A discussion shall be held if a ticket gains different labels, and the final label for the ticket could then be determined and refined through the discussion.

### 4.1 Issue Distribution

The data analysis team separately labels the attribute labels and outcome labels. According to Table 4, the most common integrity issue types that the 3 tables in (*DW-X*) have are data missing and data conflict. A smaller number of the tickets suffer from the issues of invalid data and data duplicates. Apart from integrity issues, we observed 5 data smells in the tables, and all the data smells belong to the sub-category of believability smells. These issues are confirmed not to break the database's integrity but to cause suspicions. According to the definition in section 3.1.2, consistency smells, syntactic smells and encoding smells can be avoided by setting up stricter constraints on patterns and value ranges; thus, we are not discovering these types of data smells in the tables studied.

Table 4. The distribution of data issue types on attribute dimension in the database.

| Attribute | Integrity Issues | | | | Data Smells | | | |
|---|---|---|---|---|---|---|---|---|
| | Missing | Invalid | Conflict | Duplicate | Believability | Consistency | Syntactic | Encoding |
| Quantity | 40 | 10 | 32 | 7 | 5 | 0 | 0 | 0 |

As data smells can only be finely categorized on the attribute dimension, we only discuss the distribution of integrity issues when it comes to the outcome dimension. Of the 89 integrity issues reported, 55 of the issues violate manually defined inter-column or inter-row rules, taking up over 50% portion of the total tickets.

Table 5. The distribution of data issue types on outcome dimension in the database.

| Outcome | Integrity Issues | | | |
|---|---|---|---|---|
| (Violation to) | Pattern | Range | Rule | Knowledge |
| Quantity | 3 | 13 | 55 | 18 |

Although separately labeled with our rules, no ticket label conflicts with the dimension alignment rules shown in Fig.3. The distribution of integrity issue types on the attribute and outcome dimensions is illustrated in the heatmap shown in Fig. 4. Note that data smells are not included in the figure, since they can not be mapped into the sub-categories in the outcome dimension. The most frequent violations are rule violations, mainly caused by data missing or data conflict.

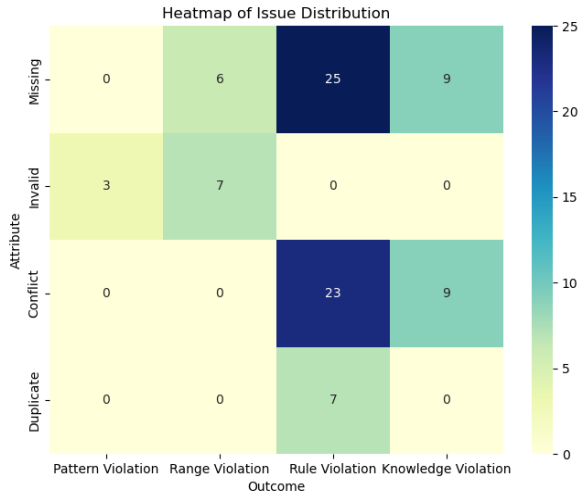

Fig. 4. The distribution of integrity issue types on attribute and outcome dimension in the database.

**Summary:** The vast majority of the reported data quality issues are integrity issues. These integrity issues can all be attributed to the outcome dimension, indicating that automated detection tools can be leveraged to identify such issues. On the other hand, our results indicate that there is a lack of attention to data smells, as we only observe a few data smells due to their latent nature. We suggest that practitioners pay more attention to such latent data issues, that future work studies their influence, and that approaches be developed to help practitioners identify such issues.

## 4.2 Issue Difficulty

The Jira issue management system offers detailed downloadable information for each reported issue. The information, including priority, severity, and time required for the fixing, can be leveraged to estimate the difficulty level of handling each type of issue. Hence, we extract *Severity, Priority, Days_to_fix, Comment_number* as four features reflecting the difficulty of issue fixing. Prior work has used similar metrics to estimate the difficulty level of issue reports (e.g., [6]) and technique forum posts (e.g., [28]). Both severity and priority are converted to integer-type labels: for priority, the level *Lowest, Low, Medium, High, Highest* are being represented by integers from 0 to 4, respectively; for severity, similarly, the level *Low, Medium, High, Critical* are being converted to integers from 0 to 3.

Table 6 depicts the four metrics' mean and max of difficulty estimations on the attribute dimension. According to the table, data duplication has the highest average severity and priority, stressing its strong impact on the data quality and downstream tasks. This type of integrity issue also has the highest average number of comments and is solved in at most 125 days, reflecting that data scientists are taking immediate action and paying much attention to data duplication issues. Data missing is another integrity issue with high average severity and priority. However, it can be noticed that data missing issues have the longest mean and maximum fixing time. Given its high importance, it can be inferred that data missing issues require long-term fixing due to the difficulty of information recovery.

Given the believability smells' legitimacy and suspicion, setting a fixed rule to eliminate these smells is difficult. According to our statistics, believability smells have a relatively high severity, but the average priority is among the lowest. They tend to have impacts on data warehouse and downstream programs but are not difficult (*i.e.,* the average days required to fix data smells is around the medium value of all the issues) to fix. A potential solution for this type of issue is to apply anomaly detection techniques [12] and let an expert define whether it is a problem. Moreover, consumers of the data should be alerted when believability smells are detected by the algorithm.

Table 6. The [mean, max] statistics for difficulty estimation with four metrics (i.e., severity, priority, days needed for fixing, and number of comments) on the attribute dimension.

| Attribute | Integrity Issues | | | | Data Smells |
|---|---|---|---|---|---|
| | Missing | Invalid | Confilct | Duplicate | Believability |
| Severity | [1.68, 3.0] | [1.2, 2.0] | [1.39, 3.0] | [2.57, 3.0] | [1.8, 2.0] |
| Priority | [2.88, 5.0] | [2.0, 3.0] | [2.41, 4.0] | [4.0, 5.0] | [1.8, 2.0] |
| Days_to_fix | [101.25, 529.0] | [97.8, 281.0] | [67.06, 378.0] | [84.29, 125.0] | [79.75, 112.0] |
| Comment_number | [15.55, 54.0] | [15.3, 37.0] | [13.53, 68.0] | [25.86, 49.0] | [15.6, 40.0] |

Table 7 reflects the metrics on the outcome dimension. As explained previously (in 4.1), we exclude statistics related to data smells in the table. Among all the violations, rule violations tend to be the severest and are being most prioritized. It can be noticed that an issue causing a rule violation can require more than a year to fix. The highest number of comments also falls into the category of rule violation, stressing its potential uncertainty in repairing. On the contrary, pattern violations have the lowest severity and priority, while all the issues categorized into this type can be quickly fixed in 63 days.

Table 7. The [mean, max] statistics for difficulty estimation with four metrics (i.e., severity, priority, days needed for fixing, and number of comments) on the outcome dimension.

| Outcome (Violation to) | Integrity Issues | | | |
|---|---|---|---|---|
| | Pattern | Range | Rule | Knowledge |
| Severity | [1.0, 1.0] | [1.46, 2.0] | [1.80, 3.0] | [1.17, 2.0] |
| Priority | [2.0, 2.0] | [2.54, 4.0] | [2.8, 5.0] | [2.61, 4.0] |
| Days_to_fix | [56.0, 63.0] | [118.46, 333.0] | [95.04, 529.0] | [46.06, 433.0] |
| Comment_number | [10.33, 18.0] | [21.92, 37.0] | [16.44, 68.0] | [9.39, 46.0] |

To gain a deeper insight into the issues, we leverage table 8 to illustrate the difficulty estimations on both the attribute and outcome dimensions. According to the table, we observed that the pattern violations reflected as data invalid have the smallest impact on the database and are easy to fix. Conversely, data duplication leading to rule violations has the highest severity and priority among all the data issues. The missing data causing rule violations takes the longest time, up to 529 days, to fix. Additionally, other data missing issues (*i.e.,* causing range violation problems or knowledge violation problems) have a high priority on average as well. These data-missing issues take a long time to fix, but data scientists tend not to have many discussions on missing data that leads to knowledge violation.

Table 8. The [mean, max] statistics on the four difficulty metrics of each type of issue.

| Outcome | Pattern Violation | Range Violation | | Rule Violation | | | Knowledge Violation | | Data Smells |
|---|---|---|---|---|---|---|---|---|---|
| Attribute | Invalid | Missing | Invalid | Missing | Conflict | Duplicate | Missing | Conflict | Believability |
| **Severity** | [1.0, 1.0] | [1.67, 2.0] | [1.29, 2.0] | [1.92, 3.0] | [1.41, 3.0] | [2.57, 3.0] | [1.0, 2.0] | [1.33, 2.0] | [1.8, 2.0] |
| **Priority** | [2.0, 2.0] | [3.17, 4.0] | [2.0, 3.0] | [2.84, 5.0] | [2.39, 4.0] | [4.0, 5.0] | [2.78, 4.0] | [2.44, 4.0] | [1.8, 2.0] |
| **Days_to_fix** | [56.0, 63.0] | [121.67, 333.0] | [115.71, 281.0] | [110.56, 529] | [81.43, 378.0] | [84.29, 125.0] | [61.78, 433] | [30.33, 150.0] | [79.75, 112.0] |
| **Comment_number** | [10.33, 18.0] | [27.17, 37.0] | [17.43, 37.0] | [16.6, 54.0] | [13.39, 68.0] | [25.86, 49.0] | [4.89, 10.0] | [13.89, 46.0] | [15.6, 40.0] |

**Summary:** We observe that rule violation issues (e.g., issues concerning multiple rows/columns) involve a relatively higher number of comments and take longer to fix, even though they are assigned a relatively higher severity and priority. One of the rule violation issues, which has the attribute of missing data, can require more than a year to repair. Our observation suggests that a detection mechanism for such violations should be in place at the data collection phase to avoid such violations from occurring.

## 5 VALIDITY THREATS

We carried out a literature review related to integrity issues and data smells. Based on the review and feedback from our industrial partner, a set of mutually exclusive taxonomies is defined on both attribute and outcome dimensions. The taxonomies are then discussed and leveraged to label real-world data quality issues from several tables in a data warehouse (*DW-X*). Validity threats to this study include internal, external, and construct validity threats.

**Internal Validity.** Our approach of leveraging the existing literature to develop the taxonomy may threaten the validity of our results: the quality of the existing taxonomies may vary. To mitigate the threat, a set of criteria is designed to ensure the quality of selected literature before the review, and the labeling rules developed are tested on tables in the real-world data warehouse, with no conflict detected.

**External Validity.** Based on our analysis, the taxonomies derived from studies on data issues in the last 10 years on outcome and attribute dimensions are mutually exclusive. The labeling rules aim to eliminate the overlap in categories and are highly generalizable to every type of database. However, due to privacy settings, only a small number of issues in several tables were reported, thus the result of data issue fixing difficulties may fail to be transferred to other cases, which could affect external validity.

**Construct Validity.** The construct validity concerns our labeling of the data quality issues. To ensure the reliability of the labels, we worked with the data analysis team from the company when labeling the reported issues in their data warehouse. We applied the method of cross-labeling to ensure the understanding of the labeling rule is aligned.

Note that our data used for labeling and repairing difficulty analysis are restricted to a small scale, due to the accessibility settings. To obtain a more comprehensive understanding of data issue fixing, more data quality issues shall be collected and analyzed.

## 6 RELATED WORKS

In this section, we summarize the literature related to the data issues addressed in our research.

**Traditional Data Issue Taxonomy.** Early data issue solutions are heavily based on rules, which require a detailed taxonomy set. Wang *et al.* categorized data quality studies into three types, namely intuitive, theoretical, and empirical [40]. Using different study strategies, data scientists have designed data quality issue taxonomies according to distinct rules: a study by Rahm *et al.* [29] defined data quality issues on the dimension of their sources, and split the categories with the existence level of the problems; Kim *et al.* [21] addressed a "near-completeness" tree-structured workflow in determining the issue types; similar to Rahm *et al.*, Oliveira *et al.* [26] split data quality issues by their sources, but eliminated the second layer of separation for definition. However, these categorizations are too fine-grained to label for academic and industry to wrangle their data quality issues (*e.g.,* the 6-layered "near-completeness" labeling strategy proposed by Kim *et al.* is not feasible for big data, which could cause excessive cost on human labor) and, thus, are not widely leveraged in recent research. To provide a more actionable and comprehensive set of taxonomy for data issue wrangling and understanding, we carried out a literature review according to a set of criteria.

**Integrity Issues.** The integrity issues, as defined in our research, are the issues that break database integrity. We collected 8 papers addressing multiple types of integrity issue-detecting systems. Abedjan *et al.* focuses on the solutions for data issues causing four types of problems: *Outliers, Duplicates, Rule Violations, and Pattern Violations* [2]; the research by Visengeriyeva *et al.* is also based on this taxonomy [36]; derived from the former research, Mahdavi *et al.* changed the category *Duplicates* to *Knowledge base violations*, bringing the taxonomies onto the dimension of outcomes, and designed a detection system Raha[25]. The other studies do not explicitly follow the previous categories, but the definitions for categories in each study share some similarities with those used in other research(*e.g.,* the *mislabels* provided by Li *et al.* [22] have the same definition with the *label errors* stated by Guha *et al.* [15]) [19] [38] [24]. However, most terminologies across these studies barely overlap, placing obstacles in issue labeling and making parallel comparisons across systems. In this work, based on the existing studies and feedback from our industry partner, we propose a comprehensive hierarchical taxonomy of data issues and study their occurrences in a real-world large-scale data warehouse.

**Data Smells.** The term "data smells" is derived from "code smells" [1] and shares many attribute similarities with code smells in the definition. A study by Sharma *et al.* [31] analyzed the root cause of data smells, and a study by Foidl *et al.* [11] defined the term based on its potential outcomes. Given the difficulty of handling and its nature of not breaking the database integrity, only a few studies focus on the topic: we collected 4 papers containing the sub-category definitions of the term. However, the sub-categories differ largely: Foidl *et al.* provided 3 simple examples to depict miscoding smells, categorical value smells and intermingled smells [11]; the research by Shome *et al.* split the categories according to issue attributes in details [33]; to refine their previous work, Foidl *et al.* classify the data smells with a highly abstracted fashion[12]; Golendukhina *et al.* equalized the dimension of *encoding smells* and *syntactic smells* in previous research with *consistency smells* and *believability smells*, and tested the credibility of the taxonomy on a real-world business travel dataset [14]. Although the sub-categories provided by Golendukhina *et al.* are well-designed and include detailed examples, data smell determination based on observability could be hard due to implicit thresholds.

## 7 CONCLUSION

In this study, we derive a comprehensive taxonomy of data quality issues from two distinct dimensions based on a literature review: the attribute dimension that indicates the intrinsic characteristics of the issues (e.g., data duplicate) and the outcome dimension that indicates the manifestation of the issues in terms of constraint violations (e.g., rule violations). Further, we bridge these two sets of taxonomies with several rules. Leveraging these rules and taxonomies, we cooperated with a real-world commercial company and labeled their data issue tickets. The labels obtained from the real-world data warehouse do not conflict with our rules, which supports the validity of our taxonomy sets and the relationships between the two dimensions.

It can be observed that most problems the data analysis team faces are rule violations, and this type of problem is also the most difficult for engineers to fix. We also observed 5 believability smells, which have moderate severity and priority but are not very hard to fix. Based on our discoveries and analyses, we provided suggestions for preventing and fixing data smells. Our comprehensive taxonomy and observations can provide guidance for practitioners and researchers to detect, classify, and fix data quality issues. However, it is worth noticing that in our case study, the dataset used for labeling is small, which may lead to a problem of generalizability in estimating the distribution of the data quality issues and the difficulties in solving them; the most impacting data quality issues may change according to domain knowledge changes. Hence, more research is required to determine the most impacting and challenging issues for a database.

## REFERENCES

[1] 1999. *Refactoring: Improving the Design of Existing Code*. Addison-Wesley Longman Publishing Co., Inc., USA.
[2] Ziawasch Abedjan, Xu Chu, Dong Deng, Raul Castro Fernandez, Ihab F Ilyas, Mourad Ouzzani, Paolo Papotti, Michael Stonebraker, and Nan Tang. 2016. Detecting data errors: Where are we and what needs to be done? *Proceedings of the VLDB Endowment* 9, 12 (2016), 993–1004.
[3] Donald P Ballou and Harold L Pazer. 1985. Modeling data and process quality in multi-input, multi-output information systems. *Management science* 31, 2 (1985), 150–162.
[4] Carlo Batini, Cinzia Cappiello, Chiara Francalanci, and Andrea Maurino. 2009. Methodologies for data quality assessment and improvement. *ACM computing surveys (CSUR)* 41, 3 (2009), 1–52.
[5] Andrew Begel and Thomas Zimmermann. 2014. Analyze this! 145 questions for data scientists in software engineering. In *Proceedings of the 36th International Conference on Software Engineering*. 12–23.
[6] Gerardo Canfora, Andrea Di Sorbo, Sara Forootani, Antonio Pirozzi, and Corrado Aaron Visaggio. 2020. Investigating the vulnerability fixing process in OSS projects: Peculiarities and challenges. *Computers & Security* 99 (2020), 102067.
[7] Tse-Hsun Chen, Weiyi Shang, Ahmed E Hassan, Mohamed Nasser, and Parminder Flora. 2016. Detecting problems in the database access code of large scale systems: An industrial experience report. In *Proceedings of the 38th International Conference on Software Engineering Companion*. 71–80.
[8] Xu Chu, John Morcos, Ihab F Ilyas, Mourad Ouzzani, Paolo Papotti, Nan Tang, and Yin Ye. 2015. Katara: A data cleaning system powered by knowledge bases and crowdsourcing. In *Proceedings of the 2015 ACM SIGMOD international conference on management of data*. 1247–1261.
[9] Roland Croft, M Ali Babar, and M Mehdi Kholoosi. 2023. Data quality for software vulnerability datasets. In *2023 IEEE/ACM 45th International Conference on Software Engineering (ICSE)*. IEEE, 121–133.
[10] Michele Dallachiesa, Amr Ebaid, Ahmed Eldawy, Ahmed Elmagarmid, Ihab F Ilyas, Mourad Ouzzani, and Nan Tang. 2013. NADEEF: a commodity data cleaning system. In *Proceedings of the 2013 ACM SIGMOD International Conference on Management of Data*. 541–552.
[11] Harald Foidl and Michael Felderer. 2019. Risk-based data validation in machine learning-based software systems. In *proceedings of the 3rd ACM SIGSOFT international workshop on machine learning techniques for software quality evaluation*. 13–18.
[12] Harald Foidl, Michael Felderer, and Rudolf Ramler. 2022. Data smells: categories, causes and consequences, and detection of suspicious data in AI-based systems. In *Proceedings of the 1st International Conference on AI Engineering: Software Engineering for AI*. 229–239.
[13] Mouzhi Ge and Markus Helfert. 2007. A Review of Information Quality Research - Develop a Research Agenda. *Proceedings of the 2007 International Conference on Information Quality, ICIQ 2007*, 76–91.
[14] Valentina Golendukhina, Harald Foidl, Michael Felderer, and Rudolf Ramler. 2022. Preliminary findings on the occurrence and causes of data smells in a real-world business travel data processing pipeline. In *Proceedings of the 2nd International Workshop on Software Engineering and AI for Data Quality in Cyber-Physical Systems/Internet of Things*. 18–21.
[15] Shubha Guha, Falaah Arif Khan, Julia Stoyanovich, and Sebastian Schelter. 2023. Automated data cleaning can hurt fairness in machine learning-based decision making. In *2023 IEEE 39th International Conference on Data Engineering (ICDE)*. IEEE, 3747–3754.
[16] Ahmed E Hassan. 2008. The road ahead for mining software repositories. In *2008 frontiers of software maintenance*. IEEE, 48–57.
[17] Anders Haug, Frederik Zachariassen, and Dennis Van Liempd. 2011. The costs of poor data quality. *Journal of Industrial Engineering and Management (JIEM)* 4, 2 (2011), 168–193.
[18] YU Huh, FR Keller, Thomas C Redman, and AR Watkins. 1990. Data quality. *Information and software technology* 32, 8 (1990), 559–565.

[19] Nick Hynes, D Sculley, and Michael Terry. 2017. The data linter: Lightweight, automated sanity checking for ml data sets. In *NIPS MLSys Workshop*, Vol. 1.
[20] Beverly K Kahn, Diane M Strong, and Richard Y Wang. 2002. Information quality benchmarks: product and service performance. *Commun. ACM* 45, 4 (2002), 184–192.
[21] Won Kim, Byoung-Ju Choi, Eui-Kyeong Hong, Soo-Kyung Kim, and Doheon Lee. 2003. A taxonomy of dirty data. *Data mining and knowledge discovery* 7 (2003), 81–99.
[22] Peng Li, Xi Rao, Jennifer Blase, Yue Zhang, Xu Chu, and Ce Zhang. 2021. CleanML: A study for evaluating the impact of data cleaning on ml classification tasks. In *2021 IEEE 37th International Conference on Data Engineering (ICDE)*. IEEE, 13–24.
[23] Gernot Liebchen and Martin Shepperd. 2016. Data sets and data quality in software engineering: Eight years on. In *Proceedings of the the 12th international conference on predictive models and data analytics in software engineering*. 1–4.
[24] Lucy Ellen Lwakatare, Ellinor Rånge, Ivica Crnkovic, and Jan Bosch. 2021. On the experiences of adopting automated data validation in an industrial machine learning project. In *2021 IEEE/ACM 43rd International Conference on Software Engineering: Software Engineering in Practice (ICSE-SEIP)*. IEEE, 248–257.
[25] Mohammad Mahdavi, Ziawasch Abedjan, Raul Castro Fernandez, Samuel Madden, Mourad Ouzzani, Michael Stonebraker, and Nan Tang. 2019. Raha: A configuration-free error detection system. In *Proceedings of the 2019 International Conference on Management of Data*. 865–882.
[26] Paulo Oliveira, Fátima Rodrigues, Pedro Henriques, and Helena Galhardas. 2005. A taxonomy of data quality problems. In *2nd Int. Workshop on Data and Information Quality*. 219–233.
[27] Charles A O'Reilly III. 1982. Variations in decision makers' use of information sources: The impact of quality and accessibility of information. *Academy of Management journal* 25, 4 (1982), 756–771.
[28] Mohamed Raed, Heng Li, Foutse Khomh, and Moses Openja. 2021. Understanding Quantum Software Engineering Challenges: An Empirical Study on Stack Exchange Forums and GitHub Issues. In *Proceedings of the 37th IEEE International Conference on Software Maintenance and Evolution (ICSME)* (2021-01-01) *(ICSME '21)*. IEEE. https://doi.org/10.1109/ICSME52107.2021.00037
[29] Erhard Rahm, Hong Hai Do, et al. 2000. Data cleaning: Problems and current approaches. *IEEE Data Eng. Bull.* 23, 4 (2000), 3–13.
[30] Thomas C Redman. 1998. The impact of poor data quality on the typical enterprise. *Commun. ACM* 41, 2 (1998), 79–82.
[31] Tushar Sharma, Marios Fragkoulis, Stamatia Rizou, Magiel Bruntink, and Diomidis Spinellis. 2018. Smelly relations: measuring and understanding database schema quality. In *Proceedings of the 40th International Conference on Software Engineering: Software Engineering in Practice*. 55–64.
[32] Martin Shepperd, Qinbao Song, Zhongbin Sun, and Carolyn Mair. 2013. Data quality: Some comments on the NASA software defect datasets. *IEEE Transactions on software engineering* 39, 9 (2013), 1208–1215.
[33] Arumoy Shome, Luis Cruz, and Arie Van Deursen. 2022. Data smells in public datasets. In *Proceedings of the 1st International Conference on AI Engineering: Software Engineering for AI*. 205–216.
[34] Antonio Torralba and Alexei A Efros. 2011. Unbiased look at dataset bias. In *CVPR 2011*. IEEE, 1521–1528.
[35] Michele Tufano, Fabio Palomba, Gabriele Bavota, Rocco Oliveto, Massimiliano Di Penta, Andrea De Lucia, and Denys Poshyvanyk. 2015. When and why your code starts to smell bad. In *2015 IEEE/ACM 37th IEEE International Conference on Software Engineering*, Vol. 1. IEEE, 403–414.
[36] Larysa Visengeriyeva and Ziawasch Abedjan. 2018. Metadata-driven error detection. In *Proceedings of the 30th International Conference on Scientific and Statistical Database Management*. 1–12.
[37] Yair Wand and Richard Y Wang. 1996. Anchoring data quality dimensions in ontological foundations. *Commun. ACM* 39, 11 (1996), 86–95.
[38] Pei Wang and Yeye He. 2019. Uni-detect: A unified approach to automated error detection in tables. In *Proceedings of the 2019 International Conference on Management of Data*. 811–828.
[39] Richard Y Wang. 1998. A product perspective on total data quality management. *Commun. ACM* 41, 2 (1998), 58–65.
[40] Richard Y Wang and Diane M Strong. 1996. Beyond accuracy: What data quality means to data consumers. *Journal of management information systems* 12, 4 (1996), 5–33.
[41] Xiaoxue Wu, Wei Zheng, Xin Xia, and David Lo. 2021. Data quality matters: A case study on data label correctness for security bug report prediction. *IEEE Transactions on Software Engineering* 48, 7 (2021), 2541–2556.
[42] Tao Xie, Suresh Thummalapenta, David Lo, and Chao Liu. 2009. Data mining for software engineering. *Computer* 42, 8 (2009), 55–62.